\documentclass[12pt]{iopart}
\usepackage{color}

\usepackage{graphicx} 

\def\beqra{\begin{eqnarray}}
\def\eeqra{\end{eqnarray}}
\def\beq{\begin{equation}}
\def\eeq{\end{equation}}
\def\ds{\displaystyle}

\def\bk{{\bf k}}
\def\vp{\varphi}

\def\bx{{\bf{x}}}
\def\bp{{\bf{p}}}
\def\bq{{\bf{q}}}

\def\agt{\stackrel{>}{\sim}}

\begin{document}

\title{Flowing with Time: a New Approach to Nonlinear Cosmological Perturbations}
\author{Massimo Pietroni}
\address{INFN, Sezione di Padova, via Marzolo 8, I-35131, Padova, Italy}
\ead{massimo.pietroni@pd.infn.it}

\begin{abstract}
Nonlinear effects are crucial in order to compute the cosmological matter power spectrum to the accuracy required by future generation surveys. Here, a new approach is presented, in which the power spectrum, the bispectrum and higher order correlations,  are obtained -- at any redshift and for any momentum scale -- by integrating a system of differential equations. The method is similar to the familiar BBGKY hierarchy. Truncating at the level of the trispectrum, the solution of the equations corresponds  to the summation of an infinite class of perturbative corrections. 
Compared to other resummation frameworks, the scheme discussed here is particularly suited to cosmologies other than $\Lambda$CDM, such as those based on modifications of gravity and those containing massive neutrinos.
As a first application, we compute the Baryonic Acoustic Oscillation feature of the power spectrum, and compare the results with perturbation theory, the halo model, and N-body simulations. The density-velocity and velocity-velocity power spectra are also computed, showing that they are much less contaminated by nonlinearities than the density-density one. 
The approach can be seen as a particular formulation of the renormalization group,  in which time is the flow parameter.
\end{abstract}

\maketitle
\section{Introduction}
Future generation galaxy surveys are going to measure the statistical properties of matter distribution to an
unprecedented accuracy, providing informations on fundamental questions such as the nature of Dark Energy (DE) \cite{EHT98,SE03} and the absolute scale of neutrino masses (see, for instance, \cite{LP06}, and references therein). 
In particular, the location and amplitude of Baryon Acoustic Oscillations 
(BAO), wiggles in the matter power-spectrum produced by the coupling 
of baryons to radiation by Thomson scattering in the early universe, 
for wavenumbers in the range $k\simeq 0.05 - 0.25 \;h \mathrm{Mpc}^{-1}$, 
have the potential to constrain the expansion 
history of the Universe and the nature of the DE. 
BAO's have recently been detected both in the 2dF and 
SDSS surveys data \cite{Eis05,Hue05,Pad06,BCB06}, and are going to be measured in the 
near future in a series of high-redshift surveys (see, for instance \cite{Hill04,GED05}).

A reliable comparison between theoretical models and observations 
requires going beyond the linear order in perturbation theory \cite{JK06,JK08,ABFL07,MP07a, MP07b,RPTBAO}. The more established way to deal with nonlinearities is by means of $N$-body simulations (for recent applications to the matter power spectrum (PS) see for instance \cite{HSW07,SSEW08,TR08}). However, in order to gain the required sensitivity, very large volumes and high resolutions are required, with the consequence that, due to time limitations, only the `vanilla' type $\Lambda$CDM cosmologies have been investigated so far. Fitting functions for the nonlinear PS, such as those based on the Halo model e.g. refs.~\cite{PD96,S03}) are uncapable to reach the required level of accuracy \cite{HSW07}.

An alternative, semi-analytical approach is represented by perturbation theory (PT) (for a review, see \cite{PT}),
which has recently experienced a renewed interest, mainly motivated by two reasons. First, next generation
galaxy surveys are going to measure the PS  at
large redshift, where the fluctuations are still in the linear regime and 
1-loop PT is expected to work \cite{JK06,JK08}. Second, techniques for the resummation of some classes of 
perturbative corrections {\em to all orders} have been recently developed. These techniques, based on field theory tools such as Feynman diagrams and the renormalization group (RG) \cite{RPTa, RPTb, MP07a, MP07b} have been shown to be able to render a PS in agreement with that from $N$-body simulations down to $z=0$ \cite{MP07a, MP07b, RPTBAO}. For related work, see \cite{Valageas03,Valageas06,McD06,Izumi07,Valageas07,Matsubara07,Taruya2007,BV08}.

While these semi-analytic models have proved themselves a viable alternative to $N$-body simulations (at least for what concerns the Dark Matter (DM) PS), they still suffer from some limitations. 
First of all, these methods are formulated for an Einstein-deSitter (EdS, $\Omega_m=1$) cosmology, and further extended to more general ones ({\it e.g.} $\Lambda$CDM) by replacing the EdS linear growth factor for the growing mode, $D^{EdS}_+=a$, with that of the other cosmology, $D_+(a)$, where $a$ is the scale factor. While this procedure is correct at the linear order, it introduces some inaccuracies at higher orders as soon as the condition $\Omega_m/f_+^2=1$ (with $f_+=d\ln D_+/d\ln a$) fails \cite{PTreview}.
The problem is that in schemes such as \cite{RPTa, RPTb, MP07a, MP07b} there is no way to independently assess the validity of such approximation. Moreover, and more importantly, such approaches cannot be trivially extended to cosmologies in which the linear growth factor is also scale dependent, $D_+=D_+(k,\,a)$, as is the case, for instance, when massive neutrinos \cite{LP06} contribute to the DM.

On a more technical ground, while the leading corrections for the PS  where correctly identified and computed in \cite{MP07a, MP07b, RPTBAO}, for the bispectrum (BS) the task turns out to be much more involved (for a computation in a different approach, see \cite{ValBisp}). Finally, a systematic scheme of approximations was not univocally identified in such frameworks, and the independent assessment of the size of the corrections one is neglecting is not a straightforward task.

In this paper we present an alternative approach to the resummation of infinite classes of PT corrections. The method is based on a straightforward application of the `equations of motion', {\it i.e.} the continuity, Euler, and Poisson equations, to the computation of correlation functions (PS, BS, etc.) for the density and the velocity fields. The correlation functions evolve in time according to a (truncated) system of differential equations, quite similar to the familiar BBGKY hierarchy ones \cite{PeebBook}.
Compared to the approaches discussed in  \cite{RPTa, RPTb, MP07a, MP07b}, the one we are going to present is much more free from field theory technicalities and, as such, will be more easily implemented by people without a specific field theoretical background. The correlation functions obey a system of integro-differential equations, 
closed by a well defined approximation procedure, whose solution gives the PS, BS, etc. at different times and for any momentum configuration. 

Remarkably, as we will discuss, the present formulation is much more flexible than the previous ones, and can be straightforwardly applied to more general cosmologies, as those based on modifications of gravity and those containing massive neutrinos \cite{LMPR08}.

As a first illustration of the possible applications of the method, we compute the PS in the BAO region. We discuss the simplest non-trivial ({\it i.e.} beyond 1-loop) approximation level and show which class of PT corrections it corresponds to. We consider two cosmologies, a `vanilla' $\Lambda$CDM, and one in which DE is a
quintessence with equation of state $w=-0.8$. For the  $\Lambda$CDM case, we compare our results with those obtained by $N$-body simulations \cite{JK06} and by other approaches, and for both cases we assess the error made by not properly treating the decaying mode. We find that it can be as large as $\simeq 1\%$ only 
for $z$ close to zero and scales smaller than those relevant for the BAO.

The density-velocity and velocity-velocity power spectra are also computed, showing that they are much less contaminated by nonlinearities than the density-density one. 

The paper is organized as follows. In Section \ref{EOM} we discuss the fluid equations for a large class of cosmologies and derive, starting from those, the evolution equations for the PS's and the BS's. We also state the only approximation made in this paper, consisting in neglecting the effect of the evolution of the connected four point functions, the trispectrum. In Section \ref{ANSOL} we derive a formal solution of the system of equations  in our approximation and show to which class of PT corrections it corresponds to. In Section \ref{COMPARE} we compare the present approach to those of refs. \cite{RPTa, RPTb, MP07a, MP07b}, and also with the one, closer in spirit, of ref.~\cite{McD06}. In Section \ref{NUMERIKA} we give and discuss our numerical results and compare it to other approaches. Finally, in Section \ref{CONCLUSION} we summarize our results and discuss future lines of work.

\section{Nonlinear fluid equations in general cosmologies}
\label{EOM}
Our starting point is given by the fluid equations,
\beqra
&&\frac{\partial\,\delta_m}{\partial\,\tau}+
{\bf \nabla}\cdot\left[(1+\delta_m) {\bf v} \right]=0\,,\nonumber\\
&& \frac{\partial\,{\bf v}}{\partial\,\tau}+{\cal H}\,({\bf v}+[{\cal A}\,{\bf v}] )\, + ( {\bf v} 
\cdot {\bf \nabla})  {\bf v}= - {\bf \nabla} \phi\,,\nonumber\\
&&\nabla^2 \phi = \frac{3}{2}\,\,{\cal H}^2  \,\Omega_m \, (\delta_m + [{\cal B}\,\delta_m] )\,,
\label{Euler}
\eeqra
which are the continuity, Euler, and Poisson equations, respectively, generalized, with respect to the EdS case, by the presence of the functions $\Omega_m(\tau)$,  ${\cal A}({\bf x},\,\tau)$,  and ${\cal B}({\bf x},\,\tau)$. We will work in conformal time, $\tau$,  and will denote by ${\cal H}= d \log a/d \tau$ the Hubble parameter,  while $\delta_m({\bf x},\,\tau)$ and  ${\bf v} ({\bf x},\,\tau)$ will be the DM number-density fluctuation and the DM peculiar velocity field, respectively.
The square brackets indicates a convolution in real space, {\it i.e.},
\beq
[\phi\,\chi]({\bf x},\,\tau)\equiv \int \frac{d^3y}{(2 \pi)^3} \,\phi({\bf x-y},\,\tau)\,\chi({\bf y},\,\tau)\,.
\eeq

The functions ${\cal A}({\bf x},\,\tau)$, ${\cal B}({\bf x},\,\tau)$, and $\Omega_m(\tau)$, parametrize different cosmologies. In the following, we list some interesting cases;
\begin{itemize}
\item[{1)}]  ${\cal A}=0$, ${\cal B}=0$, and $\Omega_m(\tau)=1$, corresponds to Einstein-de Sitter cosmology; 
\item[{2)}] ${\cal A}=0$, ${\cal B}=0$, and $\Omega_m(\tau)=\left[1+\rho_Q^0/\rho_m^0 f(\tau) \right]^{-1}$, corresponds to a flat cosmology containing DM and some non-clusterized quintessence fluid. The case of constant equation of state is given by $ f(\tau) = a(\tau)^{-3w}$, and $\Lambda$CDM corresponds to $w=-1$;
\item[{3)}] Non-vanishing values for both ${\cal A}$ and ${\cal B}$ occur when particles' geodesics are modified, as it happens for instance when the DM particles couple to some light scalar field \cite{MPS03,Mota2004,Am03} or in Scalar-Tensor theories of gravity \cite{CPS04,PMP04}. In the latter case, if $\alpha(\vp)$ represents the coupling between matter and the scalar field (in terms of the Brans-Dicke parameter $\omega$, we have $\alpha^2=1/(2 \omega+3)$) one has
${\cal A}(\bk,\,\tau) = {\cal A}(\tau)= \alpha \, d\vp/d\log\,a $ and  ${\cal B}(\bk,\,\tau)  = 2 \alpha^2 \left(1+m_\vp^2 a^2/k^2\right)^{-1}$, where $m_\vp$ is the scalar field mass, and we have transformed to Fourier space.
\end{itemize}
The approach we are going to present in this paper can be straightforwardly extended to the case of many fluctuating components. In this case, the RHS of the Poisson equation becomes
\beq 
\frac{3}{2}\,\,{\cal H}^2  \, \sum_i\Omega_i \,\delta_i\,,
\label{Rpoiss}
\eeq 
where the sum extends to all the components having non-vanishing density fluctuations. For each component,  nonlinear continuity and Euler equations should be taken into account. In some cases, however, the non-DM components have fluctuations well inside the linear regime in all scales of interest. This happens, for instance, for neutrinos and for most quintessence models. In these cases, one can approximate eq.~(\ref{Rpoiss}) 
as 
\beq 
\frac{3}{2}\,\,{\cal H}^2  \, \Omega_m \,\delta_m\left(1+ \sum_i \frac{\Omega_i \,\delta^{\rm lin}_i}{ \Omega_m \,\delta^{\rm lin}_m}\right)\,,
\label{APPLIN}
\eeq 
where the terms inside the parenthesis are taken from linear perturbation theory. Then, in this approximation, the system reduces to eqs.~(\ref{Euler}) for a single DM component with ${\cal A}=0$ and ${\cal B} (\bk, \tau) = \sum_i \Omega_i \,\delta^{\rm lin}_i/ ( \Omega_m \,\delta^{\rm lin}_m)$.

Defining, as usual, the velocity divergence $\theta(\bx,\,\tau) = 
\nabla \cdot {\bf v} (\bx, \,\tau)$, and going to
Fourier space, the equations in ~(\ref{Euler}) give
\beqra
&&\frac{\partial\,\delta_m({\bf k}, \tau)}{\partial\,\tau}+\theta({\bf k}, 
\tau) \nonumber\\
&& \qquad + \int d^3\bq\, d^3\bp \,\delta_D({\bf k}-\bq-\bp)
 \alpha(\bq,\bp)\theta(\bq, \tau)\delta_m(\bp, \tau)=0\,,\nonumber\\
&&\frac{\partial\,\theta({\bf k}, \tau)}{\partial\,\tau}+
{\cal H}\,(1+{\cal A}({\bf k}, \tau))\theta({\bf k}, \tau)  +\frac{3}{2} {\cal H}^2 \,(1+{\cal B}({\bf k}, \tau))\Omega_m(\tau)
\delta_m({\bf k}, \tau)\nonumber\\
&& \qquad   +\int d^3\bq \,d^3\bp \,\delta_D({\bf k}-\bq-\bp) 
\beta(\bq,\bp)\theta(\bq, \tau)\theta(\bp, \tau) = 0 \,.\label{EulerFourier}
\eeqra
The nonlinearity and non-locality of the Vlasov equation survive in 
the two functions
\beq\alpha(\bq,\bp )= \frac{(\bp + \bq) \cdot \bq}{q^2}\,,\quad \quad
\beta(\bq,\bp ) = \frac{(\bp + \bq)^2 \,
\bp \cdot \bq}{2 \,p ^2 q^2}\,,
\eeq
which couple different modes of density and velocity fluctuations. Notice that the functions ${\cal A}$ and 
${\cal B}$ only appear in linear terms.

One 
can write Eqs.~(\ref{EulerFourier}) in a compact form. First, 
we introduce the doublet $\vp_a$ ($a=1,2$), given by
\beq\left(\begin{array}{c}
\varphi_1 ( {\bf k}, \eta)\\
\varphi_2 ( {\bf k}, \eta)  
\end{array}\right)
\equiv 
e^{-\eta} \left( \begin{array}{c}
\delta_m  ( {\bf k}, \eta) \\
-\theta  ( {\bf k}, \eta)/{\cal H}
\end{array}
\right)\,,
\label{doppietto}
\eeq
where the time variable has been replaced by the logarithm of 
the scale factor,
\[ 
\eta= \log\frac{a}{a_{in}}\,,
\]
$a_{in}$ being the scale factor at a conveniently remote epoch, 
were all the relevant scales are well inside the linear regime. 

Then, we define a {\it vertex} function, 
$\gamma_{abc}({\bf k},{\bf p},{\bf q}) $ ($a,b,c,=1,2$) 
whose only independent, non-vanishing, elements are
\beqra
&&\gamma_{121}({\bf k},\,{\bf p},\,{\bf q}) = 
\frac{1}{2} \,\delta_D ({\bf k}+{\bf p}+{\bf q})\, 
\alpha(\bp,\bq)\,,\nonumber\\
&&\gamma_{222}({\bf k},\,{\bf p},\,{\bf q}) = 
\delta_D ({\bf k}+{\bf p}+{\bf q})\, \beta(\bp,\bq)\,,
\label{vertice}
\eeqra
and 
$\gamma_{121}({\bf k},\,{\bf p},\,{\bf q})  = 
\gamma_{112}({\bf k},\,{\bf q},\,{\bf p}) $.

The two equations (\ref{EulerFourier}) can now be rewritten as 
\beq
\partial_\eta\,\varphi_a({\bf k}, \eta)= -\Omega_{ab}({\bf k},\,\eta )
\varphi_b({\bf k}, \eta) + e^\eta 
\gamma_{abc}({\bf k},\,-{\bf p},\,-{\bf q})  
\varphi_b({\bf p}, \eta )\,\varphi_c({\bf q}, \eta ),
\label{compact}
\eeq
where 
\beq{\bf \Omega} ({\bf k},\,\eta ) = \left(\begin{array}{cc}
\ds 1 & \ds -1\\
\ds -\frac{3}{2} \Omega_m(\eta) (1+{\cal B}({\bf k},\,\eta )) & \ds 2 +
\frac{\cal H^\prime}{\cal H} +{\cal A}({\bf k},\,\eta ) \end{array}
\right)\,.
\label{bigomega}
\eeq
Repeated indices are summed over, and integration over momenta $\bq$ and $\bp$ is understood.
Notice that all the effect of changing the cosmology from EdS is contained in the matrix $\Omega_{ab}$, the vertices being universal.

The $\eta$-evolution of the correlation functions can be derived by iterating the application of eq.~(\ref{compact}),
as follows:
\beqra
 \partial_\eta\,\langle \varphi_a \varphi_b\rangle &=& -\Omega_{ac} 
\langle \varphi_c \varphi_b\rangle   - 
\Omega_{bc} 
\langle \varphi_a \varphi_c\rangle 
\nonumber\\
&&+e^\eta \gamma_{acd}\langle \varphi_c\varphi_d \varphi_b\rangle +e^\eta \gamma_{bcd}\langle \varphi_a\varphi_c \varphi_d\rangle\,,\nonumber\\
&&\nonumber\\
 \partial_\eta\,\langle \varphi_a \varphi_b  \varphi_c \rangle & =&  -\Omega_{ad} 
\langle \varphi_d \varphi_b\varphi_c\rangle  -\Omega_{bd} 
\langle \varphi_a \varphi_d\varphi_c\rangle  -\Omega_{cd} 
\langle \varphi_a \varphi_b\varphi_d\rangle
\nonumber\\
&&+e^\eta \gamma_{ade}\langle \varphi_d\varphi_e \varphi_b\varphi_c\rangle  
+e^\eta \gamma_{bde}\langle \varphi_a\varphi_d \varphi_e\varphi_c\rangle \nonumber\\
&&+e^\eta \gamma_{cde}\langle \varphi_a\varphi_b \varphi_d\varphi_e\rangle \,,
\nonumber \\
&&\nonumber\\
 \partial_\eta\,\langle \varphi_a \varphi_b  \varphi_c  \varphi_d \rangle &=& \cdots\nonumber\\
 &&\nonumber\\
& \cdots&
\label{tower}
\eeqra
In order to have more compact equations, we have omitted the momentum and $\eta$-dependence of the correlation functions. All the fields are evaluated at the same $\eta$-value. Next we express the two --three --and four-point correlators as
\beqra
&&\ds\langle \varphi_a({\bf k},\,\eta) \varphi_b({\bf q},\,\eta)\rangle \equiv \delta_D({\bf k + q}) P_{ab}({\bf k}\,,\eta)\,,\nonumber\\
&&\ds\langle \varphi_a({\bf k},\,\eta) \varphi_b({\bf q},\,\eta)\varphi_c({\bf p},\,\eta)\rangle \equiv \delta_D({\bf k + q+p})
 B_{abc}({\bf k},\,{\bf q},\,{\bf p};\,\eta)\,,\nonumber\\
 &&\ds\langle \varphi_a({\bf k},\,\eta) \varphi_b({\bf q},\,\eta)\varphi_c({\bf p},\,\eta)\varphi_d({\bf r},\,\eta)\rangle \equiv\nonumber\\
&& \ds\qquad\qquad \qquad  \left[\delta_D({\bf k + q })\, \delta_D({\bf p + r }) P_{ab}({\bf k}\,,\eta)P_{cd}({\bf p}\,,\eta)\right.\nonumber\\
&&\ds\qquad\qquad \qquad +\delta_D({\bf k + p}) \,\delta_D({\bf q + r }) P_{ac}({\bf k}\,,\eta)P_{bd}({\bf q}\,,\eta)\nonumber\\
&&\ds\qquad\qquad \qquad +\delta_D({\bf k + r})\, \delta_D({\bf q + p }) P_{ad}({\bf k}\,,\eta)P_{bc}({\bf q}\,,\eta)\nonumber\\
&& \ds\qquad\qquad \qquad \left. +\,\delta_D({\bf k + p+q+ r}) \,Q_{abcd}({\bf k}\,,{\bf q}\,,{\bf p}\,,{\bf r}\,,\eta)\right]\,,
 \eeqra
where $P_{ab}({\bf k}\,,\eta)$ is the PS, $B_{abc}({\bf k},\,{\bf q},\,{\bf p};\,\eta)$ the BS, and
$Q_{abcd}({\bf k}\,,{\bf q}\,,{\bf p}\,,{\bf r}\,,\eta)$, the connected part of the four-point function, is the trispectrum. 

In this paper, we will make the approximation $Q_{abcd} =0$. It should be emphasized that this choice, although allowing us to split the four-point functions in terms of two-point ones as in the Wick theorem, by no means amounts to consider the fields $\vp_a$ to be Gaussian, since the BS is fully taken into account. In the following Section, we will show explicitly by using diagrammatic methods, which class of non-linear interactions are kept into account by this approximation. 

The first two equations in (\ref{tower}) form now a closed system, given by:
\beqra
&&\ds  \partial_\eta\,P_{ab}({\bf k}\,,\eta) = - \Omega_{ac} ({\bf k}\,,\eta)P_{cb}({\bf k}\,,\eta)  - \Omega_{bc} ({\bf k}\,,\eta)P_{ac}({\bf k}\,,\eta) \nonumber\\
&&\qquad\qquad\quad\quad+e^\eta \int d^3 q\, \left[ \gamma_{acd}({\bf k},\,{\bf -q},\,{\bf q-k})\,B_{bcd}({\bf k},\,{\bf -q},\,{\bf q-k};\,\eta)\right.\nonumber\\
&&\qquad\qquad\qquad\qquad\qquad\left. + B_{acd}({\bf k},\,{\bf -q},\,{\bf q-k};\,\eta)\,\gamma_{bcd}({\bf k},\,{\bf -q},\,{\bf q-k})\right]\,,\nonumber\\
&&\nonumber\\
&&\ds  \partial_\eta\,B_{abc}({\bf k},\,{\bf -q},\,{\bf q-k};\,\eta) =  - \Omega_{ad} ({\bf k}\,,\eta)B_{dbc}({\bf k},\,{\bf -q},\,{\bf q-k};\,\eta)\nonumber\\
&&\qquad\qquad\qquad\qquad\qquad\quad- \Omega_{bd} ({\bf -q}\,,\eta)B_{adc}({\bf k},\,{\bf -q},\,{\bf q-k};\,\eta)\nonumber\\
&&\qquad\qquad\qquad\qquad\qquad\quad - \Omega_{cd} ({\bf q-k}\,,\eta)B_{abd}({\bf k},\,{\bf -q},\,{\bf q-k};\,\eta)\nonumber\\
&&\qquad\qquad\qquad\qquad + 2 e^\eta \left[ \gamma_{ade}({\bf k},\,{\bf -q},\,{\bf q-k}) P_{db}({\bf q}\,,\eta)P_{ec}({\bf k-q}\,,\eta)\right.\nonumber\\
&&\qquad\qquad\qquad\qquad\quad +\gamma_{bde}({\bf -q},\,{\bf q-k},\,{\bf k}) P_{dc}({\bf k-q}\,,\eta)P_{ea}({\bf k}\,,\eta)\nonumber\\
&&\qquad\qquad\qquad\qquad\quad +\left. \gamma_{cde}({\bf q-k},\,{\bf k},\,{\bf -q}) P_{da}({\bf k}\,,\eta)P_{eb}({\bf q}\,,\eta)\right]\,.
\label{syst}
\eeqra
Notice that we have restored the momentum dependences, and that the only momentum integration is the one explicitly indicated in the first of eqs.~(\ref{syst}). In Section \ref{NUMERIKA} we will present the result of the numerical integration of the equations above, without any further approximation, so, the reader interested in the performances of the present approach can skip directly to that section. In the next two sections, we discuss the relation of this method with PT, with the resummation schemes of refs \cite{RPTa, RPTb, MP07a, MP07b, RPTBAO}, and with the RG approach of ref. \cite{McD06}.
\section{Analytic Solutions}
\label{ANSOL}
The formal solution of the system (\ref{syst}) is given by
\beqra
&& P_{ab}({\bf k}\,,\eta) = g_{ac}({\bf k}\,,\eta,0)  \, g_{bd}({\bf k}\,,\eta,0)  P_{cd}({\bf k}\,,\eta=0) \nonumber\\
&&  
\qquad\qquad\quad+\int_0^\eta d\eta^\prime e^{\eta^\prime} \int d^3 q \,g_{ae}({\bf k}\,,\eta,\eta^\prime) g_{bf}({\bf k}\,,\eta,\eta^\prime) \nonumber\\
&&\qquad\qquad\qquad\quad\ \times\left[  \gamma_{ecd}({\bf k},\,{\bf -q},\,{\bf q-k})\,B_{fcd}({\bf k},\,{\bf -q},\,{\bf q-k};\,\eta^\prime)\right.\nonumber\\
&&\qquad\qquad\qquad\qquad \quad+
\left. \gamma_{fcd}({\bf k},\,{\bf -q},\,{\bf q-k})\,B_{ecd}({\bf k},\,{\bf -q},\,{\bf q-k};\,\eta^\prime)\right]\,,\nonumber\\
&&\nonumber\\
&&B_{abc}({\bf k},\,{\bf -q},\,{\bf q-k};\,\eta)=\nonumber\\
&&
 \qquad g_{ad}({\bf k}\,,\eta,0)g_{be}({\bf -q}\,,\eta,0) g_{cf}({\bf q-k}\,,\eta,0)B_{def}({\bf k},\,{\bf -q},\,{\bf q-k};\,\eta=0)\nonumber\\
&&
 \qquad\qquad\qquad +2 \int_0^\eta d\eta^\prime e^{\eta^\prime} \,g_{ad}({\bf k}\,,\eta,\eta^\prime) g_{be}({\bf -q}\,,\eta,\eta^\prime) g_{cf}({\bf q-k}\,,\eta,\eta^\prime)\nonumber\\
&&
\qquad \qquad\qquad\qquad \times\left[ \gamma_{dgh}({\bf k},\,{\bf -q},\,{\bf q-k})P_{eg}({\bf q}\,,\eta^\prime)P_{fh}({\bf q-k}\,,\eta^\prime)\right.\nonumber\\
&& \quad\qquad\qquad\qquad\qquad + \gamma_{egh}({\bf -q},\,{\bf q-k},\,{\bf k})P_{fg}({\bf q-k}\,,\eta^\prime)P_{dh}({\bf k}\,,\eta^\prime)\nonumber\\
&& \quad\qquad\qquad\qquad\qquad  \left.
+ \gamma_{fgh}({\bf q-k},\,{\bf k},\,{\bf -q})P_{dg}({\bf k}\,,\eta^\prime)P_{eh}({\bf q}\,,\eta^\prime)
\right]\,,
\label{formalsol}
\eeqra
where $g_{ab}({\bf k}\,,\eta,\eta^\prime)$ is the {\em linear propagator} \cite{RPTa, RPTb, MP07a, MP07b} which gives the $\eta$-evolution of the field at the linear level, $\varphi^L_a({\bf k}, \eta) = g_{ab}({\bf k}\,,\eta,\eta^\prime) \varphi^L_b({\bf k}, \eta^\prime)$, where the subscript ``$L$" indicates the linear approximation. In order to check the validity of the solutions above, the two following properties of the propagator should be used,
\beq
\partial_\eta g_{ab}({\bf k}\,,\eta,\eta^\prime) = -\Omega_{ac}({\bf k}\,,\eta ) g_{cb}({\bf k}\,,\eta,\eta^\prime)\,,\quad
g_{ab}({\bf k}\,,\eta,\eta) = \delta_{ab}\,.
\eeq
The explicit form of the linear propagator for the general class of cosmologies encompassed by eq.~(\ref{bigomega}) is given in \ref{approp}.

It is instructive to expand eqs.~(\ref{formalsol}) in powers of the interaction vertex $\gamma_{abc}$, in order to 
understand what class of peturbative corrections are mantained, and what are neglected, by doing the approximation $Q_{abcd}=0$.
The lowest order, corresponding to linear perturbation theory, is obtained by setting $\gamma_{abc}=0$ on the RHS of eqs.~(\ref{formalsol}), yelding,
\beqra
&& P^L_{ab}({\bf k}\,,\eta) = g_{ac}({\bf k}\,,\eta,0)  \, g_{bd}({\bf k}\,,\eta,0)  P_{cd}({\bf k}\,,\eta=0)\,, \nonumber\\
&&\nonumber\\
&&B^L_{abc}({\bf k},\,{\bf -q},\,{\bf q-k};\,\eta)=\nonumber\\
&&
 g_{ad}({\bf k}\,,\eta,0)g_{be}({\bf -q}\,,\eta,0) g_{cf}({\bf q-k}\,,\eta,0)B_{def}({\bf k},\,{\bf -q},\,{\bf q-k};\,\eta=0)
 \,,
 \eeqra
 where we explicitly see, besides the initial PS, the initial BS evaluated at $\eta=0$. Notice that, due to the $\e^{-\eta}$ factor in front of eq.~(\ref{doppietto}), the linear PS is independent on $\eta$, on the growing mode. 
 
\begin{figure}
\centerline{\includegraphics[width = 2.5in,keepaspectratio=true]{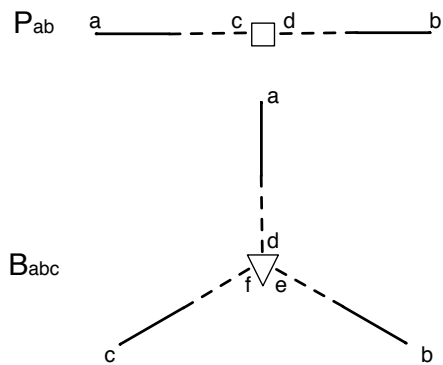}}
\caption{The $O(\gamma^0)$ contributions to the power spectrum and the bispectrum.}
\label{FIGURALINEARE}
\end{figure}

\begin{figure}
\centerline{\includegraphics[width = 4.in,keepaspectratio=true]{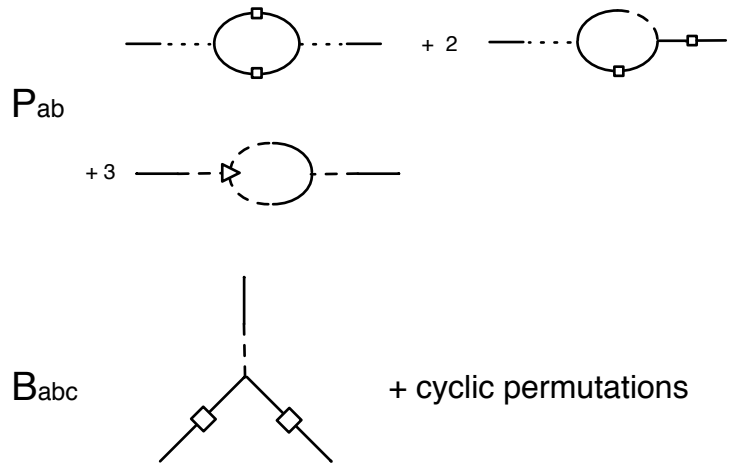}}
\caption{The $O(\gamma)$ contributions to the bispectrum and the $O(\gamma^2)$ (first line)   and $O(\gamma)$ (second line) ones to power spectrum.}
\label{FIG1LOOP}
\end{figure}

\begin{figure}
\centerline{\includegraphics[width = 3.5in,keepaspectratio=true]{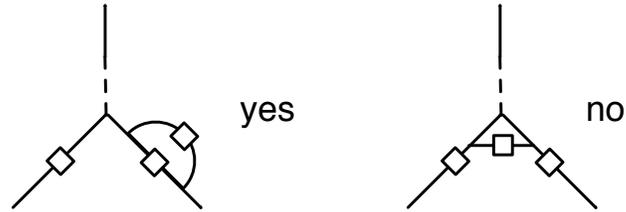}}
\caption{$O(\gamma^3)$ corrections to the BS included (left) and not included (right) in the approximation scheme considered in this paper, namely $Q_{abcd}=0$.}
\label{BCORR}
\end{figure}

 In the language of Feynman diagrams introduced in \cite{MP07b}, the linear order solution can be represented as in fig.~\ref{FIGURALINEARE}, where the lines correspond to propagators, the empty box to the primordial PS and the empty triangle to the primordial BS.
The next order ($O(\gamma)$) correction for the BS is obtained by inserting $ P^L_{ab}$ in place of $P_{ab}$ at the RHS of the second of eqs.~(\ref{formalsol}).  The result is represented at the bottom of fig.~\ref{FIG1LOOP}, where the vertex represent the interaction $\gamma_{abc}$. Inserting the BS at this order (that is, the sum of the  $O(\gamma^0)$  and $O(\gamma)$ contributions of figs.~\ref{FIGURALINEARE} and \ref{FIG1LOOP}, respectively) in the first of eqs.~(\ref{formalsol}), we get the $O(\gamma^2)$ and $O(\gamma)$ contributions to the PS, represented in the first two lines of fig.~\ref{FIG1LOOP}. At this order, the result for the PS coincides with the  standard 1-loop expression in eulerian perturbation theory \cite{PT}, as can be checked by using the composition rule for the propagators,
\beq
 g_{ab}({\bf k}\,,\eta,\eta^\prime) g_{bc}({\bf k}\,,\eta^\prime,\eta^{\prime\prime})
= g_{ac}({\bf k}\,,\eta,\eta^{\prime\prime})\,.
 \eeq
Iterating the procedure one more time, the first differences with respect to perturbation theory arise. Considering 
the BS, $O(\gamma^3)$ and $O(\gamma^2)$ corrections are generated by substituting one of the two power spectra lines in fig.~\ref{FIG1LOOP} with the $O(\gamma^2)$ or  $O(\gamma)$ contributions to the PS, one possibility being represented on the left of fig.~\ref{BCORR}. However, in perturbation theory, there are other  $O(\gamma^3)$ corrections which cannot be obtained in this way, such as that on the right of fig.~\ref{BCORR}.  In a field theoretic language, as a consequence of  the approximation $Q_{abcd}=0$,  we are neglecting vertex renormalization, while including the renormalization of the PS. This finding generalizes at any order. The $n$-th iteration leads to contributions which are given by sums of terms having the same structure as fig.~\ref{FIG1LOOP}, with the legs containing a box replaced $O(n-1)$ contributions to the PS, while the vertices, propagators, and triangles are left untouched. This is well summarized by the expressions in eq.~(\ref{formalsol}), where the PS is formally 1-loop (there is a momentum integration, $d^3 q$), while the BS is formally tree-level (no momentum integration).

The results presented in Sect.\ref{NUMERIKA} correspond to the resummation of this class of contributions to all order in $\gamma$.

\section{Comparison to other approaches}
\label{COMPARE}
From the discussion of the previous section, one can realize that the results of linear perturbation theory for the PS are recovered by setting $\gamma_{abc}$ to zero on the RHS of the integro-differential equations in ~(\ref{syst}). The 1-loop approximation corresponds instead to setting $P_{ab}({ \bf p},\eta) =  P_{ab}({ \bf p},\eta=0)$ on the RHS of the second of eqs.~(\ref{syst}), while keeping the first one in its full form. 

Finally, keeping the $\eta$-dependence of the PS at the RHS of both equations corresponds to resumming, to all orders, the
perturbative corrections in which the interaction vertex is kept at its tree-level form, $\gamma_{abc}$, {\it i.e.} in which no diagrams like the one at the right of fig.~\ref{BCORR} are included. This class of corrections is the same which was considered in refs.~\cite{MP07b}, where RG equations were also solved under the approximation of keeping the tree-level expression for the vertex. Indeed, the results of the present paper compare very well with those obtained via the RG. However, there are also some practical differences between the two approaches, which are worth while mentioning. 

First of all, in the RG case, as well as in the renormalized perturbation theory of Crocce and Scoccimarro \cite{RPTa, RPTb}, the computation of the PS requires two steps, each one implying further approximations 
besides the one of keeping the vertex at tree-level. In the first step, the propagator is computed. As was shown in \cite{RPTa, RPTb, MP07a, MP07b} a simple analytic expression, exact in the limit of large external momentum, can be obtained for the propagator at any order in perturbation theory. However, the use of this large momentum limit turns out to give inaccurate results \cite{RPTBAO}, so that subleading corrections have to be taken into account. In the second step, the `resummed' propagator is used to compute the PS. The calculation is performed either  by making extra approximations to the relevant RG equations \cite{MP07b}, or at finite loop order, as in \cite{RPTBAO}. 

The remarkable feature of the present approach is that there is no need to compute the propagator in order to get the PS (or the BS, etc.). The information on the time evolution, which, in the approaches of refs.~\cite{RPTa, RPTb, MP07a, MP07b, RPTBAO} is contained in the propagator, here is treated exactly by the differential (in time) structure of the equations. Once the $Q_{abcd}=0$ truncation is performed, the system is solved without any further approximation. Also, the next level of approximation is clearly identified in keeping $Q_{abcd}$ and truncating at the five-point correlator level, and so on.

Moreover,  as we already discussed in the Introduction, using differential equations in time has the other precious virtue of allowing a clear treatment of the general class of cosmologies described by eqs.~\ref{Euler}. Strictly speaking, approaches such as ~\cite{MP07a, MP07b, RPTBAO} are exact only in the Einstein-deSitter case ($\Omega_M=1$), in which the matrix ${\bf \Omega} ({\bf k},\,\eta )$ appearing in the the equation of motion reads
\beq{\bf \Omega} = \left(\begin{array}{cc}
\ds 1 & \ds -1\\
\ds -\frac{3}{2}  &\ds \frac{3}{2} \end{array}
\right)\,.
\label{omegaEds}
\eeq
In cosmologies such as $\Lambda$CDM, where $\Omega_m < 1$ and the linear growing mode for matter, $D_+$, is scale-independent, if one makes the following replacements
\beq
\eta \rightarrow \log \frac{D_+(a)}{D_+(a_{in})}\,,\qquad\varphi_2({\bf k},\eta) \rightarrow \frac{\varphi_2({\bf k},\eta)}{f_+(\eta)}\,,
\label{changevar}
\eeq
with $f_+= d\log D_+/d\log a$,
one gets equations of the same form as eq.(\ref{compact}) but with 
\beq{\bf \Omega} (\eta ) = \left(\begin{array}{cc}
\ds 1 & \ds -1\\
\ds -\frac{3 \Omega_m}{2f_+^2}  &\ds \frac{3 \Omega_m}{2f_+^2} \end{array}
\right)\,.
\label{omegappr}
\eeq
The phenomenologically interesting cases of $\Lambda$CDM and (non-clusterizing) quintessence are then treated in ~\cite{RPTa, RPTb, MP07a, MP07b, RPTBAO} by using eq.~(\ref{changevar}) and approximating eq.~(\ref{omegappr}) with (\ref{omegaEds}). It is easy to realize (see \ref{approp}) that in this approximation the growing mode of matter perturbations is treated properly, whereas the decaying mode has the wrong time dependence and the wrong ratio between density and velocity perturbations. Since the vertex $\gamma_{abc}$ mixes growing and decay modes, this approximation is expected to work well in the linear regime, and to fail when nonlinearities become important\footnote{Another way of seeing the problem \cite{PTreview} is to notice that if $\Omega_m/f_+^2\neq 1$ the equations of motion (\ref{compact}) do not admit separable solutions of the form $\varphi_a(\bk,\,\eta) = \sum_n D_+^n(\eta) \chi_a(\bk)$. }. The frameworks discussed in ~\cite{RPTa,RPTb, MP07a, MP07b, RPTBAO} have no way of assessing the validity of the approximation, while, in the present approach, one can directly test it by comparing the results obtained with the constant  and the  $\eta$- dependent ${\bf \Omega}$ matrices. In the next section we will discuss such a comparison (see fig. \ref{Diff}).
The most generic case of scale dependent growth factors is precluded to the approaches such as ~\cite{RPTa, RPTb, MP07a, MP07b, RPTBAO}, while it can be properly treated here. The most relevant example of massive neutrinos will be the subject of a future publication \cite{LMPR08}.

Evolution equations for the PS depending on time (or, better, on the growth factor), were discussed by Mc Donald in \cite{McD06}. They are equivalent (at least in the approximated approach to $\Lambda$CDM discussed above) to make the approximation 
\beq
P_{ab}({\bf p},\,\eta) \simeq u_a \,u_b\;P({\bf p},\,\eta) \,,\qquad u = \left(\begin{array}{c} 1\\1\end{array}\right)\,,
\label{macapp}
\eeq 
in the last of eqs.(\ref{formalsol}), and in inserting the BS obtained in this way into the first of eqs.(\ref{syst}). By doing so, one obtains an evolution equation whose RHS has the structure of the 1-loop correction to the PS, in which the linear ($\eta$-independent) PS has been replaced by the renormalized ($\eta$-dependent) one. As we will see in the next section (see figs.~\ref{Pij_z1}, \ref{Pij_z0}), the different components of the PS exhibit sizably different evolutions, therefore one expects that the approximation in  eq.~(\ref{macapp}) would lead to results far from the accuracy required in applications such as the BAO, as it seems to be the case of ref.~\cite{McD06}.
\begin{figure}
\centerline{\includegraphics[width = 5.in,keepaspectratio=true]{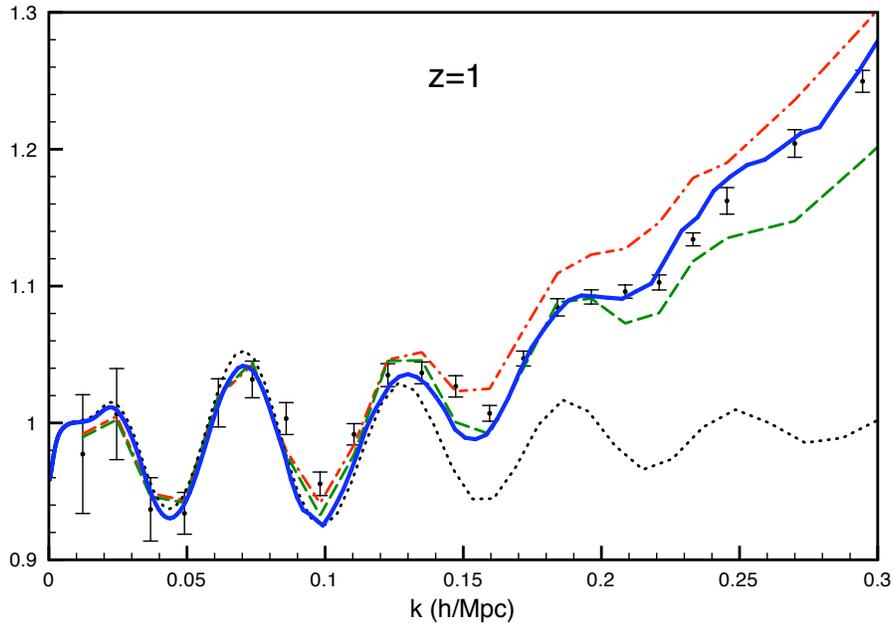}}
\caption{Power spectra at redshift $z=1$ (divided by a smooth one). The continuous line is the result of the present paper, compared with linear theory (dotted), 1-loop PT (dash-dotted), the halo approach of ref.~\cite{S03} (dashed). The dots with error bars are taken from the $N$-body simulationd of ref.~\cite{JK06}. The background cosmology is a spatially flat $\Lambda$CDM model with $\Omega_\Lambda^0=0.73$, $\Omega_b^0=0.043$, $h=0.7$, $n_s=1$, $\sigma_8 = 0.8$.}
\label{OMZ_z1}
\end{figure}

\begin{figure}
\centerline{\includegraphics[width = 5.in,keepaspectratio=true]{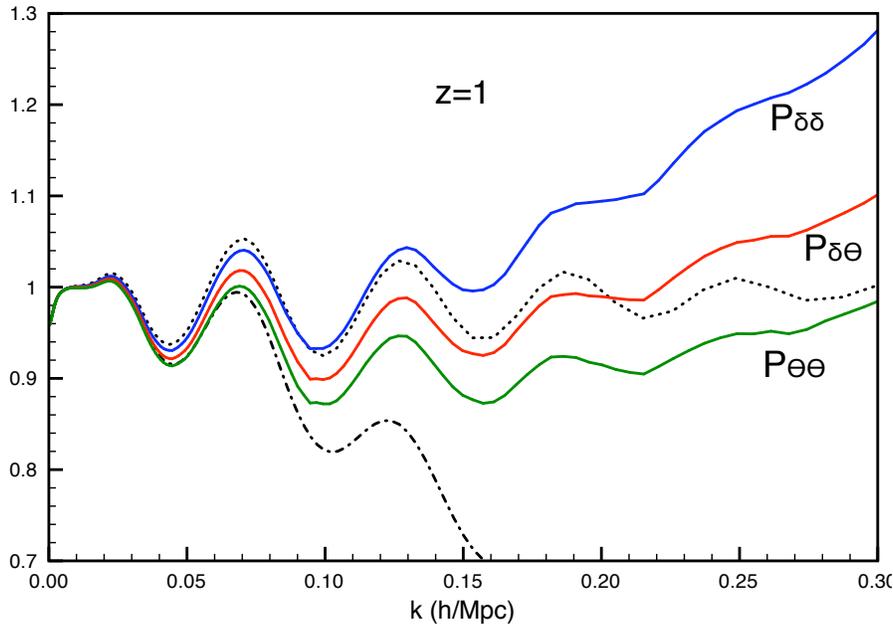}}
\caption{The density-density ($P_{\delta\delta}$), density-velocity ($P_{\delta\theta}$), and velocity-velocity ($P_{\theta\theta}$) power spectra at $z=1$. The dotted line is linear theory, and the dash-dotted one the linear theory multiplied by $\exp(-k^2 \sigma_v^2(e^\eta-1)^2)$. }
\label{Pij_z1}
\end{figure}
\begin{figure}
\centerline{\includegraphics[width = 5.in,keepaspectratio=true]{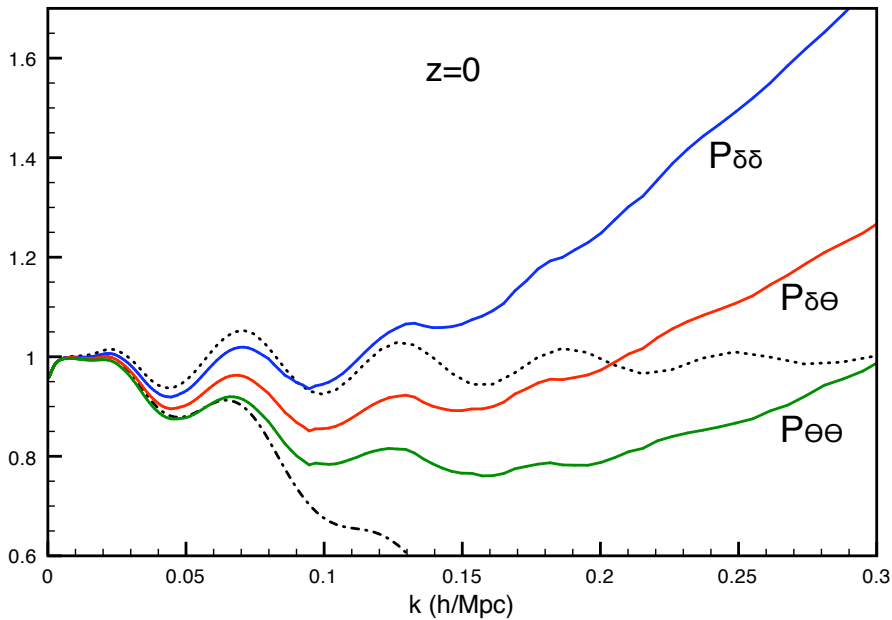}}
\caption{Same as fig.~\ref{Pij_z1}, but at $z=0$.}
\label{Pij_z0}
\end{figure}
\section{Numerical Results}
\label{NUMERIKA}
In this section, we present the results of a numerical integration of the equations (\ref{syst}). Our code is built along the lines described in \ref{numdet}.
As benchmark cosmologies, we will consider a flat $\Lambda$CDM model, with  $\Omega_\Lambda^0=0.73$, $\Omega_b^0=0.043$, $h=0.7$, $n_s=1$, $\sigma_8 = 0.8$, and a quintessence model with equation of state $w=-0.8$ and the other parameters set at the same values. The linear growth factors are scale-independent in both cases, so we will perform the change of variables in eq.~(\ref{changevar}) and use the exact --$\eta$-dependent -- matrix ${\bf \Omega}$ in eq.~(\ref{omegappr}). 

The integration starts at a $\eta_{in}=0$ corresponding to a high redshift. In this paper, we set $z_{in}=35$. The initial conditions at $\eta=0$ are the linear PS, as obtained by running the CAMB code \cite{CAMB}, and vanishing BS, {\it i.e.}, we are ignoring all non-gaussianities generated at higher redshifts. Then, the evolution equations are evolved down to low redshift, giving the PS and the BS for any momentum scale. In fig.\ref{OMZ_z1} we give the (density-density) PS for our $\Lambda$CDM cosmology, divided by a smooth spectrum, obtained by setting $\Omega_b^0=0$. The results of this paper are represented by the continuous line. For comparison, we also give the results of linear theory (dotted) of 1-loop PT, dash-dotted, of the halo approach of ref.~\cite{S03} (dashed), and of the $N$-body simulations of ref.~\cite{JK06}.
Compared to the RG approach of refs.\cite{MP07a, MP07b}, the agreement with $N$-body simulations is equally good in the BAO region, but here, it extends to higher $k$'s. 

In figs.~\ref{Pij_z1} and \ref{Pij_z0} we plot the density-density, density-velocity, and the velocity-velocity PS, given by $P_{11}$, $P_{12}$, and $P_{22}$, respectively. As it was discussed in \cite{MP07a, MP07b, RPTBAO}, the exact PS has the structure
\beq
P_{ab}(k,\,\eta) = G_{ac}(k,\eta,0)G_{bd}(k,\eta,0)P_{cd}(k,\, 0) + P^{II}_{ab}(k,\,\eta) \,
\label{somma}
\eeq
where $G_{ac}(k,\eta,0)$ is the full propagator. The first term on the RHS can be approximated as \cite{RPTBAO}
\beq
e^{(-k^2 \sigma_v^2(e^\eta-1)^2)} P^L(k,\, 0)\,
\eeq
with $P^L(k,\, 0)$ the linear PS, and 
\beq
\sigma_v^2 = \frac{1}{3}\int d^3 q \frac{ P^L(q,\, 0)}{q^2}\,,
\eeq
the velocity dispersion. This contribution is also plotted in the figures, with a dash-dotted line. In this approximation, the first contribution to eq.~(\ref{somma}) is the same for the three components of the PS. The difference between them is then entirely due to the mode-coupling contribution $P^{II}_{ab}(k,\,\eta)$, which has the effect of smoothing out the BAO feature \cite{RPTBAO}. As we see from the figures, the velocity-velocity and the density-velocity PS are less affected by $P^{II}$ than the density-density one, therefore they should be taken into serious consideration as alternative keys to measuring the BAOs. 

\begin{figure}
\centerline{\includegraphics[width = 5.in,keepaspectratio=true]{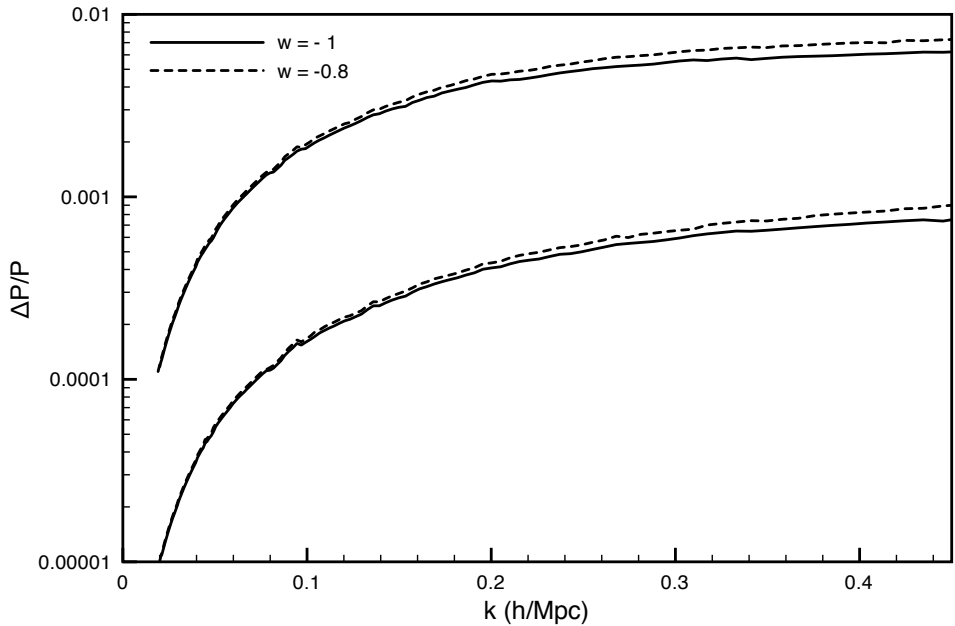}}
\caption{Relative difference between the power spectrum computed with the time-dependent matrix ${\bf \Omega}$ in eq.~(\ref{omegappr}) and the approximated, constant one, eq.~(\ref{omegaEds}). Continuous lines are for $\Lambda$CDM ($w=-1$), and dashed ones for a quintessence with constant equation of state $w=-0.8$. For each cosmology, the bottom (top) lines correspond to $z=1$ ($z=0$).}
\label{Diff}
\end{figure}

Finally, we address the issue of the error made by approximating the exact ${\bf \Omega}$ in eq.~(\ref{omegappr})  with the constant one in eq.(\ref{omegaEds}), as is usually done in PT as well as in the approaches of refs.\cite{RPTa, RPTb, MP07a, MP07b, RPTBAO}. This approximation should become less and less accurate at decreasing redshifts. Also, it should have some dependence  on the cosmology under consideration, since the ratio $\Omega_m/f_+^2$ increases from 1 at high redshift to 1.157 at redshift zero for $\Lambda$CDM and to 1.179 for the $w=-0.8$ quintessence cosmology. In fig.~\ref{Diff} we plot the relative difference between the PS computed with the time-dependent matrix ${\bf \Omega}$ and the approximated, constant one. Continuous lines are for $\Lambda$CDM ($w=-1$), and dashed ones for a quintessence with constant equation of state $w=-0.8$. For each cosmology, the lines from bottom to top correspond to $z=1,\,0$, respectively. We see that, in both cases, the relative error in the BAO range of stays below the per mill  level at $z =1$, and below the percent at $z=0$. Therefore, for high redshift surveys, our results indicate that the approximation is well motivated.

\section{Summary}
\label{CONCLUSION}
Compared to previous semi-analytic approaches to the resummation of perturbative corrections, the scheme presented in this paper has some clear advantages. First of all, it can be easily extended to a large class of  cosmologies, among which the case of massive neutrinos and of scalar-tensor modifications of gravity are the most notable ones. In practice, the approach in its present form can be used for all models in which the nonlinear terms in the continuity and in the Euler equations are the same as for the EdS case. The evolution can be seen as a series of time steps (in the limit of vanishing step length), during which the fields evolve with the --possibly scale-dependent-- linear growth factors. At the end of each step, the growing and decaying mode are mixed by the universal nonlinear terms. Therefore, the different cosmologies enter only via the different background evolution and the different linear growth factor, all the information being contained in the ${\bf \Omega}$ matrix of eq.~(\ref{bigomega}).

The second advantage is the possibility of formulating a clean and systematic scheme of approximations. The next level, compared to the one considered here, is obviously given by the inclusion of the trispectrum in the system of equations, which will correspond to including vertex corrections. Judging from the present agreement between our results and $N$-body simulations in the BAO region, one would infer that such improved approximation will mainly affect higher momentum scales.

For a more practical point of view, though the present approach is completely compatible with those of refs.~\cite{RPTa,RPTb,MP07a,MP07b} (see the discussion in Section \ref{COMPARE}), here field theory tools such as Feynman diagrams, generating functionals, and RG flows, are not necessary in order to compute, for instance, the PS and the BS. What is needed, at the end, is just a code to solve the differential equations in eq.(\ref{syst}). Therefore, a field theoretic background is not a prerequisite to use this method.

The results we obtain are in good agreement with the N-body simulations of ref. \cite{JK06}, up to $k$ of order $0.3 - 0.35 \,\mathrm{h/Mpc}$, showing that the method is able to describe (from first principles) the non-linearities due to mode-mode coupling in the BAO region. We also showed that the velocity-density and the velocity-velocity PS's are less contaminated by the mode-mode coupling and, since they should also be much less contaminated by bias, their possible role in the future BAO measurements should be studied in more detail.

Finally, we showed that the role of the decaying mode can be safely ignored for $z \agt1$, and can lead to percent-level effects only at lower redshifts and scales higher than those relevant for the BAO's.

The importance of nonlinear effects for the determination of the upper bound on the neutrino mass scale was emphasized in ref. \cite{STT08}, where 1-loop PT was used to compute the PS. A better nonlinear approach is clearly desirable to derive firmer neutrino mass bounds \cite{LMPR08}, possibly including the running of the trispectrum, in order to reach higher momentum scales.

\appendix
\section{The linear propagator in general cosmologies} 
\label{approp}
The linear equation
\beq
\partial_\eta\,\varphi_a({\bf k}, \eta)= -\Omega_{ab}({\bf k},\,\eta )
\varphi_b({\bf k}, \eta) ,
\label{linear}
\eeq
has solutions of the form
\beq
\left(\begin{array}{c}1\\ f({\bf k}, \eta)
\end{array}\right)\,\varphi({\bf k}, \eta) ,
\label{growdec}
\eeq
where $\varphi$ and $f$ satisfy, 
\beqra
&&\partial_\eta\, \varphi =- (\Omega_{11} +\Omega_{12} f  ) \varphi  \nonumber\\
&&\partial_\eta\, f = \Omega_{12}  f^2 +(\Omega_{11}-\Omega_{22})  f -\Omega_{21} \nonumber\\ 
&&\quad \;\;\;=\Omega_{12} (f-\bar{f}_+)(f-\bar{f}_-)\,,
\label{equaz}
\eeqra
with
\beq
\bar{f}_\pm({\bf k}, \eta) = \frac{\Omega_{22}-\Omega_{11}\mp\sqrt{(\Omega_{22}-\Omega_{11})^2+4\Omega_{12}\Omega_{21}}}{2\Omega_{12}}\,.
\label{fbar}
\eeq
Solving the equations (\ref{equaz}), we get the evolution in $\eta$ of the upper and lower components of (\ref{growdec}), 
\beqra
&&\ds \varphi(\eta) = \e^{-\int_{\eta^\prime}^\eta dx (\Omega_{11} +\Omega_{12} f)} \varphi(\eta^\prime)\,,\nonumber\\
&&\nonumber\\
&& \ds f(\eta) \varphi(\eta) = \e^{-\int_{\eta^\prime}^\eta dx (\Omega_{21}/  f+\Omega_{22})} 
 f(\eta^\prime) \varphi(\eta^\prime) \nonumber\\
 &&\qquad\qquad\;\;\,= \e^{-\int_{\eta^\prime}^\eta dx (\Omega_{11} +\Omega_{12} f)} \frac{ f(\eta)}{ f(\eta^\prime)}\,
 f(\eta^\prime) \varphi(\eta^\prime)\,.
\eeqra
Next, we identify two independent solutions of eq.~(\ref{equaz}), which will serve as basis for the expansion of the most generic solution of eq.~(\ref{linear}). In most cases, it is convenient to choose as basis solutions the growing and decaying modes. For instance, if the Universe at high redshift tends to Einstein-deSitter, one can identify the basis solutions by setting their initial conditions as
\beq
f_\pm({\bf k}, \eta_i) = \bar{f}_\pm({\bf k}, \eta_i)\,,
\eeq
at some early time $\eta_i$, with $ \bar{f}_\pm$ given in eq.~(\ref{fbar}).

We define the instantaneous projectors on the two basis solutions, for each $\eta \ge \eta_i$,
\beqra
&&{\bf M}^+({\bf k}, \eta) \left(\begin{array}{c}1\\ f_{+}({\bf k}, \eta)
\end{array}\right) = \left(\begin{array}{c}1\\ f_{+}({\bf k}, \eta)
\end{array}\right)\,,\nonumber\\
&&{\bf M}^+({\bf k}, \eta) \left(\begin{array}{c}1\\ f_{-}({\bf k}, \eta)
\end{array}\right) =0\,,
\eeqra
which is given explicitely by
\beq
{\bf M}^+ = \frac{1}{f_- -f_+}\left(\begin{array}{cc}
f_- & -1\\
f_- f_+ & -f_+
\end{array}
\right)\,.
\eeq
The projector on the other mode is given by ${\bf M}^- = {\bf 1} - {\bf M}^+$. 

The propagator, defined as the operator giving the linear evolution of the field $\varphi_a$, {\it i.e}, 
$\varphi_a({\bf k}, \eta) = g_{ab}({\bf k}, \eta, \eta^\prime ) \varphi_b({\bf k}, \eta^\prime)$, is then given by
\beqra
&& {\bf g}({\bf k}, \eta, \eta^\prime ) = \e^{-\int_{\eta^\prime}^\eta dx (\Omega_{11} +\Omega_{12} f_+)} 
\left(\begin{array}{cc}
1 & 0\\
0 & \frac{f_+({\bf k}, \eta)}{f_+({\bf k}, \eta^\prime)}
 \end{array} \right) {\bf M}^+({\bf k}, \eta^\prime) \nonumber\\
 &&\qquad\qquad\quad +  \e^{-\int_{\eta^\prime}^\eta dx (\Omega_{11} +\Omega_{12} f_-)} 
\left(\begin{array}{cc}
1 & 0\\
0 & \frac{f_-({\bf k}, \eta)}{f_-({\bf k}, \eta^\prime)}
 \end{array} \right) {\bf M}^-({\bf k}, \eta^\prime)\,,
 \eeqra
 for $\eta>\eta^\prime$, and zero otherwise.
 One can check that it satisfies the properties
  \beqra
  && \partial_\eta\, {\bf g} ({\bf k}\,,\eta,\eta^\prime) = -{\bf \Omega} ({\bf k}\,,\eta ) \cdot {\bf g}({\bf k}\,,\eta,\eta^\prime)\,,\nonumber\\
  && \lim_{\eta\prime->\eta^{-}} {\bf g}({\bf k}, \eta, \eta^\prime ) = {\bf 1}\,,\nonumber\\
  && {\bf g}({\bf k}, \eta, \eta^\prime ) \cdot {\bf g}({\bf k}, \eta^\prime, \eta^{\prime\prime } ) = {\bf g}({\bf k}, \eta, \eta^{\prime\prime })\,,
\eeqra
and that, in the Einstein-deSitter case ($\Omega_{11}=-\Omega_{12}=1$, $\Omega_{22}=-\Omega_{21}= \frac{3}{2}$), it reduces to the propagator considered in \cite{MP07b}, with $f_+=1$, $f_-=-3/2$.

\section{Putting the equations in a more numerical-friendly form} 
\label{numdet}
In case of $k$-independent (but $\eta$-dependent) $\Omega$ matrices, the evolution equation for the PS, eq.~(\ref{syst}) can be rewritten as
\beqra
&&\ds  \partial_\eta\,P_{ab}(k) = - \Omega_{ac} P_{cb}(k)  - \Omega_{bc} P_{ac}( k) \nonumber\\
&&\qquad\qquad\quad\quad+ e^\eta \frac{4 \pi}{k} \int_{k/2}^\infty dq \,q \int_{|q-k|}^q dp \,p\, \left[ \tilde{\gamma}_{acd}(k, q, p)\, \tilde{B}_{bcd}(k,q,p)\right.\nonumber\\
&&\qquad\qquad\qquad\qquad\qquad\qquad\qquad\qquad\left. + \tilde{B}_{acd}(k,q,p)\,\tilde{\gamma}_{bcd}(k,q,p)\right]\,,
\label{runP}
\eeqra
where the $\eta$-dependence of the PS, the BS and the ${\bf \Omega}$ matrix is understood, and we have defined
\beq
 \tilde{\gamma}_{abc}(k, q, p) = \left. \gamma_{abc}({\bf k}, {\bf q}, {\bf p})\right|_{{\bf p}=-({\bf k}+{\bf q})}\,,
\eeq
and analogously for $\tilde{B}_{abc}(k, q, p)$. In numerical applications, the integrals appearing in the above expression are  quite time consuming to perform, both because the integration interval for the internal variable ($p$) depends on the external one ($q$), and because the integrand contain a function, 
$\tilde{B}_{bef}(k,q,p)$, which has to be computed via an independent evolution equation, eq.~(\ref{syst}), for each set of its three variables. 
In order to put the equations in a more manageable form, it is convenient to define the following quantities,
\beqra 
&& I_{acd,bef}(k) \equiv \int_{k/2}^\infty dq \,q \int_{|q-k|}^q dp \,p\, \frac{1}{2} \left[ \tilde{\gamma}_{acd}(k, q, p)\, \tilde{B}_{bef}(k,q,p) + (q\leftrightarrow p) \right].\nonumber\\
&&
\eeqra
Then, eq.~(\ref{runP}) can be written as
\beqra
&&\ds  \partial_\eta\,P_{ab}(k) = - \Omega_{ac} P_{cb}(k)  - \Omega_{bc} P_{ac}( k) \nonumber\\
&&\qquad\qquad\quad + e^\eta \frac{4 \pi}{k} \left[I_{acd,bcd}(k)+I_{bcd,acd}(k) \right]\,.
\label{runfig}
\eeqra
The evolution of the $I_{acd,bef}(k)$ can be deduced from that for the bispectra, {\it i.e.} from the second of  eqs.~(\ref{syst}), and is given by
\beqra
&&\ds  \partial_\eta\,I_{acd,bef}(k) =- \Omega_{bg} I_{acd,gef}(k) - \Omega_{eg}  I_{acd,bgf}(k) \nonumber \\
&& 
\qquad\qquad\qquad - \Omega_{fg} I_{acd,beg}(k) +2 e^\eta A_{acd,bef}(k)\,,
\label{runI}
\eeqra
with the integrals $A_{acd,bef}(k)$ given by
\beqra
&&A_{acd,bef}(k) \equiv  \int_{k/2}^\infty dq \,q \int_{|q-k|}^q dp \,p\, \frac{1}{2} \left\{ \tilde{\gamma}_{acd}(k, q, p) \left[ \tilde{\gamma}_{bgh}(k, q, p)P_{ge}(q) P_{hf}(p)\right.\right. 
\nonumber\\
&& 
\left.\left. + \,\tilde{\gamma}_{egh}(q,p,k)P_{gf}(p) P_{hb}(k) +\tilde{\gamma}_{fgh}(p,k,q)P_{gb}(k) P_{he}(q)\right] +
(q\leftrightarrow p)\right\}\,.
\label{Aint}
\eeqra
Comparing (\ref{runP})  to (\ref{Aint}), we see that the latter integral contains only running functions of one variable, the $P$'s.

The evolution of the PS can then be obtained by solving the system of eqs.~(\ref{runfig}) and (\ref{runI}). By inspection of the explicit expression for the vertex, eq.~(\ref{vertice}), and of the symmetry properties of the integral (\ref{Aint}), one can conclude that  of the 64 components of $I_{acd,bef}(k)$ only 12 take part to the system. Indeed, the 12 independent components to follow are identified by the direct product of the two 
triplets $(acd)=(112),\,(222)$ and the six triplets $(bef)=(b11),\,(b12),\,(b22)$, ($b=1,2$). Together with the 3 equations for the independent components of the PS (namely, $P_{11}$, $P_{12}$, and $P_{22}$), we have a total of 15 equations.

It is also convenient to put the integrals such as eq.~(\ref{Aint}) in the following form,
\beqra
&&\int_{k/2}^\infty dq \,q \int_{|q-k|}^q dp \,p\,  \left[F(k,q,p) + (q\leftrightarrow p) \right]\nonumber\\
&& \qquad =
\int_{k/\sqrt{2}}^\infty dx \int_0^{k/\sqrt{2}} dy\,\frac{x^2-y^2}{2}\left[F(k,\frac{x+y}{\sqrt{2}},\frac{x-y}{\sqrt{2}}) + (y\leftrightarrow -y) \right],
\eeqra
which has the virtue of having both integration intervals independent on the integration variables.
\noindent
\section*{Acknowledgments}
It is a pleasure to thank Guido D'Amico, Julien Lesgourgues, Sabino Matarrese and Toni Riotto for many enlightening discussions.
\section*{References}
\bibliographystyle{JHEP}
\bibliography{mybib}
\end{document}